# Stochastic Optimal Control in Continuous Space-Time Multi-Agent Systems

Wim Wiegerinck    Bart van den Broek    Bert Kappen
SNN, Radboud University Nijmegen
6525 EZ Nijmegen, The Netherlands
{w.wiegerinck,b.vandenbroek,b.kappen}@science.ru.nl

## Abstract

Recently, a theory for stochastic optimal control in non-linear dynamical systems in continuous space-time has been developed (Kappen, 2005). We apply this theory to collaborative multi-agent systems. The agents evolve according to a given non-linear dynamics with additive Wiener noise. Each agent can control its own dynamics. The goal is to minimize the accumulated joint cost, which consists of a state dependent term and a term that is quadratic in the control. We focus on systems of non-interacting agents that have to distribute themselves optimally over a number of targets, given a set of end-costs for the different possible agent-target combinations. We show that optimal control is the combinatorial sum of independent single-agent single-target optimal controls weighted by a factor proportional to the end-costs of the different combinations. Thus, multi-agent control is related to a standard graphical model inference problem. The additional computational cost compared to single-agent control is exponential in the tree-width of the graph specifying the combinatorial sum times the number of targets. We illustrate the result by simulations of systems with up to 42 agents.

## 1 INTRODUCTION

A collaborative multi-agent system (MAS) is a collection of agents that autonomously control their behavior to achieve a common goal or to maximize the performance of the group. Examples are teams of soccer-robots and teams of unmanned rescue vehicles in a hazardous disaster area. In practical applications, a MAS often has to deal with uncertainty in the environment and limitations of its resources.

In this paper, we are interested in optimal control in such systems. We focus on systems in which agents in a stochastic environment have to distribute themselves efficiently over a number of targets. For example, consider a system of $n$ firefighter-agents and fires. The agents are at some initial positions and should reach the fires positions in the most efficient way, such that each fire is reached by an agent (see figure 1). In this problem, the final configuration, i.e., which agent has reached exactly which fire is not of importance for the end performance. The MAS should continuously control itself such that in the end one of these $n!$ configurations is realized at minimal expected effort. The additional complexity is that due to the noise in the dynamics, a configuration that seems optimal from the initial positions may become suboptimal in a later stage.

A common approach is to model such a system as a Markov Decision Process (MDP) in discrete space and time: the optimal actions in an MDP optimization problem are in principle solved by backward dynamic programming. Since both the joint action space and the joint state space of the agents are assumed to be large in the discretization, and increase exponentially in the number of agents, simply taking a basic dynamic programming approach to solve the MDP will generally be infeasible [1].

Typically one can describe the system more compactly as a factored MDP. In such systems both the transition probabilities and reward functions have some structure. Unfortunately, this structure is not conserved in the value functions and exact computation remains exponential in the system size. Recently, a number of advanced and powerful approximate methods have been proposed. The common denominator of these approaches is that they basically assume some predefined approximate structure of the value functions [2, 3].

In this paper, we take a different starting point.

Rather than discretizing, we will consider the stochastic optimal control problem in continuous space and time. As in discrete MDPs, this optimization problem is in principle solved by backward dynamic programming. Usually this optimization is intractable. However, if (1) both the noise and the control are additive to the (nonlinear) dynamics, (2) the increment in cost is quadratic in the control, and (3) the noise satisfies certain additional conditions, then it can be shown that the stochastic optimization problem can be transformed into a linear partial differential equation, which can be solved by forward stochastic integration of a diffusion process [4, 5]. This formalism contains linear-quadratic control as a special case [6].

An interesting observation in [4, 5] is the phenomenon of symmetry breaking in multi-modal systems, i.e, in problems where several local minima of the cost coexist. This symmetry breaking manifests itself as a delayed choice, keeping options open and using the fact that the noise may help to come closer to one of the options at no additional cost.

Formally the extension of this formalism to cooperative MAS is straightforward. The question that we ask ourselves, is how the formalism scales from the single-agent single-target situation (e.g. one fireman has to go to a given fire) to a collaborative system of $n$ agents and $m$ targets. Although the dynamics of the agents is assumed to be independent, with optimal control the behavior of the agents will be coupled in a non-trivial way in order to reach an end-configuration at minimal cost.

The paper is organized as follows. In section 2 we provide a review of the general (single-agent) framework. As an example, we will rederive linear quadratic control from control theory. In this system a single agent has to control itself to a single target. Next, in section 3 we show how the framework easily allows the modeling of an agent that has to control itself to one of $m$ possible targets. It turns out that optimal control for this case can be written as a weighted sum of $m$ optimal controls in the presence of the single targets. This result will form the basis of the multi-agent analysis later in the paper.

In section 4 we will consider the framework in the multi-agent setting. In general, the solution of this type of problem consists of a sum of $m^n$ terms due to a contribution from each agent-target combination. For small problems, this summation can be performed explicitly. For large $n$ and $m$, the solution is generally intractable.

Next, we consider models in which the end-costs of agents-targets configurations are factored as a sparse graph. We show that the structure of the graph is con-served in the $m^n$ terms. Therefore this summation can be performed more efficiently using e.g. the junction tree algorithm [7]. The computation time is exponential in the induced tree-width of the graph times the number of targets $m$, and is linear in the number of agents $n$. This is in contrast to discrete MDPs, where, as remarked earlier, sparsity in the reward function is not retained in the value function and does not help to reduce the computation costs.

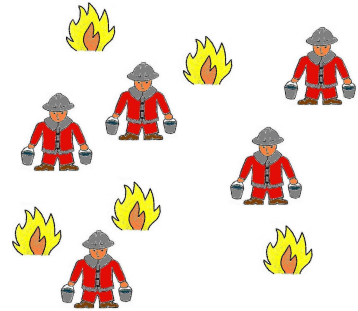

Figure 1: The firemen problem. A number of firemen go to a number of fires, each to a different one. How should the agents coordinate when it is not yet decided to which fire each agent should go, and when the actions of the agents are subject to noise?

In section 5, we illustrate the framework by simulation results of stochastic optimal control in the two toy problems considered in this paper.

## 2 STOCHASTIC OPTIMAL CONTROL OF A SINGLE AGENT

In this section, we review the framework developed in [4, 5].

We consider an agent moving in $\mathbb{R}^k$. Its position $x$ obeys the stochastic dynamics

$$dx = (b(x,t) + u)dt + d\xi, \qquad (1)$$

with $d\xi$ a Wiener process with $\langle d\xi_i d\xi_j \rangle = \nu_{ij} dt$, $b(x,t)$ an arbitrary function of $x$ and $t$, modeling the dynamics due to the environment. The agent can influence the dynamics by the control $u$.

Given $x$ at initial time $t_i$, the problem is to find a control trajectory $u(t_i \to T)$ such that the expected cost-to-go

$$C(x, t_i, u(t_i \to T)) =$$
$$\left\langle \phi(x(T)) + \int_{t_i}^T dt \left( \frac{1}{2} u(t)^\top R u(t) + V(x(t), t) \right) \right\rangle \quad (2)$$

is minimal. The expectation is taken over all noise realizations, resulting in different trajectories in state space $x(t_i \to T)$ that start in $x$. $\phi(x(T))$ is the end cost, depending only on the end state $x(T)$. $V(x(t),t)dt$ is the cost of being at position $x(t)$ during the time interval $[t, t+dt]$, $dt\, u(t)^\top R u(t)$ is the cost of the control during the same time interval. $R$ is a constant $k \times k$ matrix.

The expected cost-to-go at time $t$ needs to be minimized over all strategies $u(t \to T)$, this yields the optimal (expected) cost-to-go

$$J(x,t) = \min_{u(t \to T)} C(x,t,u(t \to T)). \tag{3}$$

In the appendix, it is briefly explained that due to the linear-quadratic form of the optimization problem—the dynamics (1) is linear in the action $u$, the cost (2) is quadratic in the action—the minimization can be performed explicitly, yielding a non-linear partial differential equation in $J$. If, in addition, the matrices $\nu$ and $R$ can be linked via a scalar $\lambda$ such that $\nu = \lambda R^{-1}$, the optimal cost-to-go is re-expressed as the log of an ordinary integral (equation (15) in the appendix),

$$J(x,t) = -\lambda \log Z(x,t) \tag{4}$$

with "partition function"

$$Z(x,t) = \int dy\, \rho(y,T|x,t) e^{-\phi(y)/\lambda} \tag{5}$$

in which $\rho(y,T|x,t)$ is the probability of arriving in state $y$ at time $T$, when starting in $x$ at time $t$, under the dynamics (14) in the appendix. This dynamics(14) equals the stochastic system dynamics without control, i.e., $u = 0$, with in addition the probability $\frac{V(x,t)}{\lambda} dt$ of being removed from the system between $t$ and $t + dt$ and thus not arriving in state $y$. $\phi(y)$ is the end cost in the state $y$.

The optimal control of the agent is directly obtained from the optimal cost-to-go, by taking its gradient (equation (10) in the appendix), which implies the following result

$$u(x,t) = \nu \partial_x \log Z(x,t). \tag{6}$$

**Example 1** *The running example is a system with linear dynamics, $b = 0$, and zero potential $V(x,t) = 0$. $R$ and $\nu$ are proportional to the identity and are considered as scalars. Regardless the end costs, the diffusion process results in a Gaussian distribution*

$$\rho(y,T|x,t) \propto \exp\left[-\frac{|y-x|^2}{2\nu(T-t)}\right].$$

*If we take a quadratic end cost around a target $\mu$,*

$$\phi(x) = \frac{\alpha}{2}|x-\mu|^2$$

*the cost to go follows from a Gaussian convolution with $e^{-\phi/\lambda}$, resulting in*

$$Z(x,t) \propto \exp\left[-\frac{|x-\mu|^2}{2\nu(T-t+R/\alpha)}\right].$$

*The optimal control follows from (6) and (5) resulting in*

$$u(x,t) = \frac{\mu - x}{T - t + R/\alpha}.$$

*This result is well known from control theory [6].*

**Example 2** *Now $b$ and $V$ are arbitrary, such that the diffusion process results in a distribution $\rho(y,T|x,t)$. To enforce an end state close to target $\mu$, a quadratic end cost with a large $\alpha$ can be chosen. The effect is that the factor $e^{-\phi(y)/\lambda}$ becomes proportional to a delta-function centered around $\mu$, and $Z$ follows directly from the value of $\rho$ at the target,*

$$Z(x,t) \propto \int dy\, \rho(y,T|x,t) \delta(y-\mu) \propto \rho(\mu,T|x,t)$$

*from which $J$ and $u$ follow directly.*

## 3 MULTIPLE TARGETS

In this section, we apply the framework of the previous section to the case where an agent has to reach one of a number of possible end states. We show that the optimal control can be constructed from a weighted combination of single-target optimal controls.

When the agent has to reach one of a number of states $\mu_1, \ldots, \mu_m$ at the end time, this can be modeled by letting $e^{-\phi(y)/\lambda}$ be a linear combination of functions which each are peaked around a single target $\mu_s$ (with $s = 1, \ldots, m$), like e.g. a delta function $\delta(y - \mu_s)$ or a Gaussian centered around $\mu_s$ with small width. We denote these functions as $\Phi(y; \mu_s) \equiv \Phi(y; s)$. If we put an additional cost $E(s)$ when target $\mu_s$ is reached, this combination becomes

$$e^{-\phi(y)/\lambda} = \sum_{s=1}^{m} e^{-E(s)/\lambda} \Phi(y;s) \equiv \sum_{s=1}^{m} w(s) \Phi(y;s),$$

Substitution into (5) gives the partition function

$$Z(x,t) = \sum_{s=1}^{m} w(s) Z(x,t;s)$$

which is a weighted combination of single-target partition function

$$Z(x,t;s) = \int dy\, \rho(y,T|x,t) \Phi(y,s).$$

The optimal action is obtained from (6) and reads

$$u(x,t) = \sum_{s=1}^{m} p(s|x,t) u(x,t;s),$$

with single-target optimal controls

$$u(x,t;s) = \nu \partial_x \log Z(x,t;s)$$

and $p(s|x,t)$ the probability

$$p(s|x,t) = \frac{w(s) Z(x,t;s)}{\sum_{s'=1}^{m} w(s') Z(x,t;s')}.$$

**Example 3** *In the running example, optimal control with multiple targets is*

$$u(x,t) = \frac{\bar{\mu} - x}{T - t + R/\alpha}$$

*with $\bar{\mu}$ the 'expected target'*

$$\bar{\mu} = \sum_{s=1}^{m} p(s|x,t) \mu_s$$

*which is the expected value of the target according to the probability*

$$p(s|x,t) = \frac{w(s) \exp\left[-\frac{|x-\mu_s|^2}{2\nu(T-t+R/\alpha)}\right]}{\sum_{s=1}^{m} w(s) \exp\left[-\frac{|x-\mu_s|^2}{2\nu(T-t+R/\alpha)}\right]}$$

## 4 STOCHASTIC OPTIMAL CONTROL OF A MAS

We now turn to the issue of optimally coordinating a multi-agent system of $n$ agents. In principle, a multi-agent system can be considered as a system with a joint state $x = (x_1, \ldots, x_n)$, where $x_a$ is the state of agent $a$, a joint dynamics (1), and a joint cost (2) which is to be minimized by a joint action $u = (u_1, \ldots, u_n)$, where $u_a$ is the control of agent $a$. The optimal control by agent $a$ follows from the appropriate components of the gradient

$$u_a(x_1, \ldots, x_n, t) = \nu \partial_{x_a} \log Z(x_1, \ldots, x_n, t). \quad (7)$$

We remark that in continuous space-time, the optimal controls can be determined independently for each agent, and coordination does not have to be imposed explicitly. This is in contrast to discrete multi-agent MDP models, in which coordination may be needed since more than one optimal joint action can exist [1]. The reason is that in continuous time control results in actions that are infinitesimal within an infinitesimal time increment. This allow agents to adapt their control immediately to each other.

### 4.1 INDEPENDENT DYNAMICS, JOINT TASK

In the remainder of the paper, we consider agents with independent dynamics $b_a(x,t) = b_a(x_a,t)$ and independent noise $\nu_{ab} = \nu_a \delta_{ab}$ with $\nu_a$ a noise matrix restricted to the domain of agent $a$. We also assume individual contributions to the costs during the process: $R_{ab} = R_a \delta_{ab}$ with $R_a$ a matrix restricted to $a$, and $V(x,t) = \sum_a V_a(x_a,t)$. We finally assume that $\nu = \lambda R^{-1}$ holds globally. Under these assumptions, the agents behave like 'non-interacting particles', e.g., they can freely move through each other without costs for collisions. Under these assumptions, the joint solution of the diffusion process factorizes into a product of solutions of independent (single agent) diffusion processes.

$$\rho(y,T|x,t) = \prod_a \rho_a(y_a,T|x_a,t) .$$

The agents optimal control and the resulting dynamics will be coupled by their joint task, expressed in the end-costs $\phi(y)$. We consider the problem where the agents have to distribute themselves over a number of targets $\mu = \mu_1, \ldots, \mu_m$.

The trivial case where each agent $a$ has to go to a single target $\mu_{s_a}$ is equivalent with a single control problem, with joint target $\mu_s = (\mu_{s_1}, \ldots, \mu_{s_n})$. Now $s$ is the vector of labels, $s = (s_1, \ldots, s_n)$. Of course, control by the agents is independent of each other. The partition function factorizes in single agent partition functions

$$\begin{aligned} Z(x,t;s) &= \prod_a \int dy_a \rho_a(y_a,T|x_a,t) \Phi_a(y_a;s_a) \\ &\equiv \prod_a Z_a(x_a,t;s_a) . \end{aligned}$$

More interesting is the case where the system has more choices in how to distribute itself. Like in the single-agent case, this is described by defining $e^{-\phi(y)/\lambda}$ to be a positive linear combination of peaked (multi-agent) single-target functions, $\Phi(y;s) = \prod_a \Phi_a(y_a;s_a)$, as in section (3), with the difference that in this sum $s$ runs over $m^n$ states (for all the possible distributions of agents over targets). The partition function of this system then reads

$$Z(x,t) = \sum_s w(s) \prod_a Z_a(x_a,t;s_a) .$$

The optimal control of an individual agent $a$ is obtained using (7), and leads again to an average of single-target optimal controls,

$$u_a(x,t) = \sum_{s_a=1}^{m} p(s_a|x,t) u_a(x_a,t;s_a),$$

where $u_a(x_a, t; s_a) = \nu_a \partial_{x_a} Z_a(x_a, t; s_a)$ and $p(s_a|x,t)$ is the probability

$$p(s_a|x,t) = \frac{\sum_{s \setminus s_a} w(s) \prod_b Z_b(x_b, t; s_b)}{\sum_s w(s) \prod_b Z_a(x_b, t; s_b)},$$

which can be interpreted as the probability that agent $a$ has to go to target $s_a$ given the joint state $x$ of the MAS and time $t$.

**Example 4** *The firemen problem. We consider $n$ identical agents, and $m$ targets modeled as in examples 1 and 3. The aim of the agents is to distribute themselves with minimal action such that each target is reached at time $T$ by about $\frac{n}{m}$ agents. We model the system by an additional cost of*

$$E(s) = c \sum_{f=1}^m \left( \sum_{a=1}^n \delta_{f,s_a} - \frac{n}{m} \right)^2 = c \left( \sum_{a,b=1}^n \delta_{s_a,s_b} - \frac{n^2}{m} \right)$$

*in which $c > 0$ is a constant indicating the costs of suboptimal distributions. Optimal control of agent $a$ is given by*

$$u_a(x,t) = \frac{\bar{\mu}_a - x_a}{T - t + R/\alpha} \qquad (8)$$

*with $\bar{\mu}$ the expected target for agent $a$,*

$$\bar{\mu}_a = \sum_{s_a} p(s_a|x,t) \mu_{s_a},$$

*where*

$$p(s_a|x,t) \propto \sum_{s \setminus s_a} \exp\left[ -\frac{E(s)}{\lambda} - \frac{|x - \mu_s|^2}{2\nu(T-t+R/\alpha)} \right]. \qquad (9)$$

The additional computation effort in multi-agent control compared to single agent control is the computation of $p(s_a|x,t)$, which involves a sum over $m^n$ states. For small systems this is feasible. For large systems, this is only feasible if the summation can be performed efficiently.

### 4.2 FACTORED END-COSTS

The issue is the computational complexity of the probability distribution

$$p(s|x,t) = \frac{1}{Z(x,t)} w(s) \prod_a Z_a(x_a, t; s_a).$$

The complexity comes from the weights $w(s)$ of the end costs, which couples the agents. In the case that the weights $w(s)$ allow a factored representation

$$w(s) = \prod_\alpha w_\alpha(s_\alpha) = \exp\left[ \sum_\alpha -\frac{E_\alpha(s_\alpha)}{\lambda} \right]$$

in which $\alpha$ are subsets of agents, we can apply the junction tree algorithm from probabilistic graphical models to make inference more efficient [7]. The complexity is then exponential in the induced tree-width of the underlying graph. In the case of the firemen problem, this approach does not really help. The clusters $\alpha$ contain only two agents. However all pairs of agents appear in a factor, which makes the graph fully connected, similar to the fully connected Boltzmann machine. The tree-width of the graph is $n$. Non-trivial tree width (smaller than $n$) can be expected in systems where the contribution of an agent to the end cost only depends on the states of a limited number of other agents.

**Example 5** *Holiday resort problem. We consider a set of $n$ agents with identical dynamics, and $m$ targets (a few holiday resorts) modeled as in examples 1 and 3. Each agent has relations with only a few other agents, and only cares for related agents whether or not to have holiday in the same resort (depending on the sign of the relation). If two agents are unrelated they are indifferent whether they end up in the same resort. Relations are assumed to be symmetric. The aim of the agents is that they have optimally distributed themselves with minimal effort over the resorts at time $T$. A way to model such a system is to define for each related pair of agents $a$ and $b$ a cost*

$$E_{ab}(s_a, s_b) = -c_{ab} \delta_{s_a, s_b}$$

*with $c_{ab}$ to weight the relevance of the relation. The sign of $c_{ab}$ is equal to the sign of the relation. The optimal control is again as in (8), with the current $E(s)$ substituted into (9). Note that the firemen problem is a special case, with a fully connected network of negative relations of strength $c$.*

Inference of graphical models is in general linear in the number of variables $n$ and exponential in the number of states in the cliques of the underlying junction tree [7]. The number of states in the largest clique is equal to the treewidth of the graph times the number of states per node. This implies that models with sparse graphs and a limited number of targets are tractable.

## 5 NUMERICAL EXAMPLES

In this section we illustrate the optimally controlled stochastic MASs by numeric simulations results.

In all simulations, we modeled agents in 1-d (for exposure purposes). The models were as in the running examples, with $b = 0$ and $V = 0$. In numerical simulations, time is to be discretized. This has to be done with a bit of care. In continuous time, $udt$ should be infinitesimal, regardless the size of $u$. In the discrete approximation this implies that $u\Delta t$ should be small compared to typical scales in the system. For the running examples, in particular (8), this means that

$$u\Delta t = \epsilon(\bar{\mu} - x)$$

with a small $\epsilon \ll 1$. From (8) we solve

$$\Delta t = \epsilon(T - t + R/\alpha)$$

which yields a finite discretization for finite $\alpha$. In the simulations, we took $\alpha = 10^3$ and $\epsilon = 0.01$. Furthermore we took noise parameter $\nu = 1$, and $R = 1$. We started all agents at $t = 0$ at $x = 0$. End time is $T = 1$.

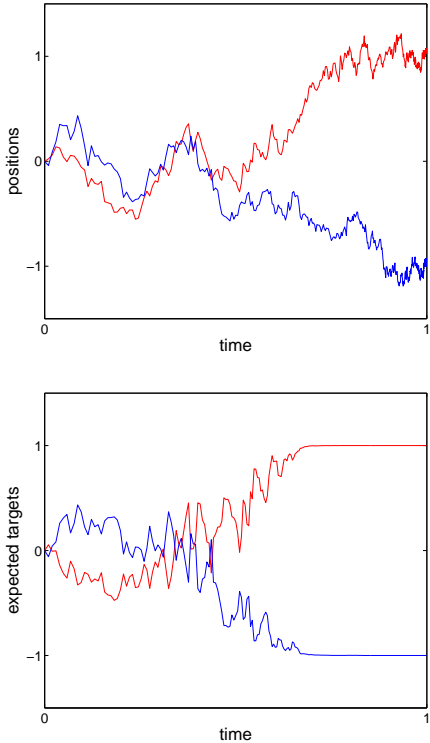

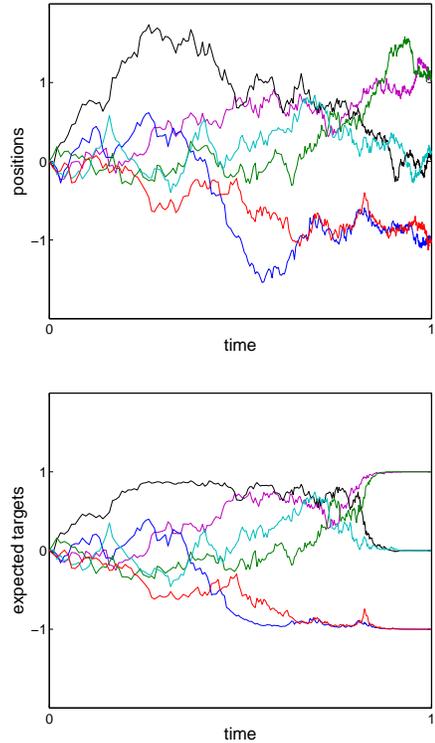

Figure 2: Simulation of firemen problem. Two agents start in $x=0$ at $t=0$ and should reach at $t=1$ the two targets located at $x=-1$ and $x=1$ in a noisy environment with minimal cost of control. (a) The positions $x_a$ of the agents as a function of time. (b) The expected targets $\bar{\mu}_a$ of the agents as a function of time.

Figure 3: Simulation of firemen problem with 6 agents start in $x=0$ at $t=0$ and should reach at $t=1$ the three targets located at $x=-1,0,1$, preferably two agents per target. (a) The positions $x_a$ of the agents as a function of time. (b) The expected targets $\bar{\mu}_a$ of the agents as a function of time.

In the first simulation, we have the firemen problem with two agents and two fires located at $-1$ and $1$. We model a preference of one agent per fire in the end situation. This is achieved by using the weight representation $w(s_1, s_2) = \exp(-E(s_1, s_2)/\lambda)$ and setting $E(1, 2) = E(2, 1) = 0$ and $E(1, 1) = E(2, 2) = 2$. In figure 2, the positions of the 2 agents $x_1$ and $x_2$ are plotted, as well as the expected target locations $\bar{\mu}_1$ and

$\bar{\mu}_2$. We see that the MAS reached a preferred goal: at the end time each target is reached by exactly one agent. During the whole trajectory, $\bar{\mu}_1 \approx -\bar{\mu}_2$ since the MAS mostly aim at an end-configuration with one agent per fire. Furthermore, note that in the first part of the trajectory, the expected targets are close to zero, while only after about $t = 0.6$ the agents seem to make a clear choice for their targets. This delayed choice is due to a symmetry breaking in the cost-to-go as time increases. Before the symmetry breaking, it is better to keep options open, and see what the effect of the noise is. After the symmetry breaking, time is too short to wait longer and a choice has to be made. This phenomenon is typical for multi-modal problems. For more details we refer to [4, 5].

In the second simulation, we have the firemen problem with six agents and three fires located at $-1$, $0$ and $1$. We modeled the end cost as in example 4. In figure 3, the positions of the 3 agents are plotted, as well as the expected target locations. From the figure it can be concluded that the MAS has successfully dis-tributed itself with two agents at each target. In this simulation, two (local) symmetry breakings are clearly visible. At about t=0.5, 2 agents seem to choose for target $\mu = -1$ and the other four agents for an expected target of $\bar{\mu} = 0.5$. Then at about $t = 0.8$ there is a second symmetry breaking, where these four agents make their final choice.

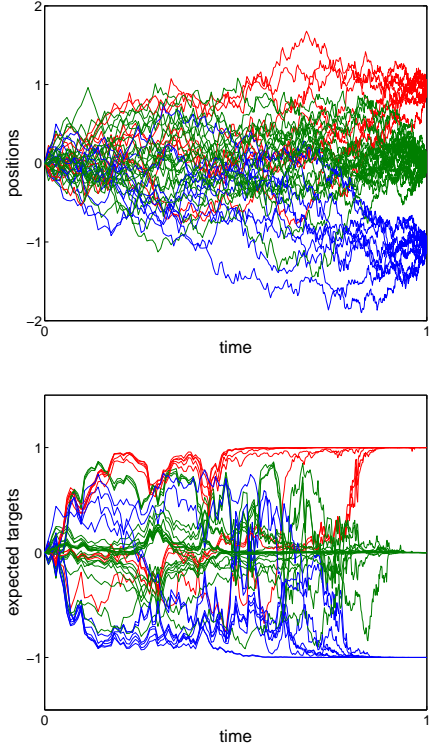

Figure 4: Simulation of holiday resort problem. 42 agents start in $x = 0$ at $t = 0$ and should reach at $t = 1$ the three targets located at $x = -1, 0, 1$, preferably together with positively related agents and not with negatively related agents (relations are not shown in the figure). (a) The positions $x_a$ of the agents as a function of time. (b) The expected targets $\overline{\mu}_a$ of the agents as a function of time.

In the last simulation, we have the holiday resort problem with 42 agents and three resorts located at $-1$, 0 and 1. We modeled the end cost as in example 4. To model the relations between the agents we generated a random graph with 42 nodes, in which each node is coupled to exactly three randomly chosen neighbors. The relation strengths were randomly chosen $c_{ab} = \pm 1$ with equal probability.

In figure 4, the positions of the 42 agents are plotted, as well as the expected target locations. From the results it can be seen that each agent reached a target. Actually, target $-1$ is reached by 10 agents, target 0 by 23 agents, and target 1 is reached by 9 agents. In this simulation, the coordination in terms of cluster formation in $\bar{\mu}$ is profound, despite the fact that the positions of the agents seem to be quite chaotic.

In the graph in this simulation, there were 34 positive and 29 negative relations. The treewidth is 7. Among the agents that ended at target $-1$, there were 5 positive relations and 0 negative ones. At target 0, there were 18 positive relations and 1 negative one. At target 1, there were 6 positive relations and 0 negative ones. So within the targets, there were a total of 29 positive relations and 1 negative one. Between agents at different targets, there were 5 positive relations and 28 negative ones.

## 6 DISCUSSION

We studied optimal control in collaborative multi-agent systems in continuous space-time. A straightforward approach to discretize the system in space and time would make the $n$ agent MAS intractable due to the exponential blow-up of the state-space. In this paper, we took the approach developed in [4, 5]. We showed that under given model assumptions, optimal distributed control can be solved analytically and that this solution is tractable in large sparsely connected systems.

In dense multi-agent systems, however, the exact inference is intractable. A natural approach would be the use of message passing algorithms for approximate inference. This is currently studied and will be reported in near future.

There are many possible model extensions that need to be explored in future research. Obvious extensions, such as a non-fixed end-time, or systems with more realistic environments, such as allowing for obstacles are already of interest to study in the single agent situation. Others apply typically to the multi-agent situation, such as penalties for collisions between agents. Typically, these types of model extensions will prohibit an analytical solution of the control, and approximate numerical methods will be required. Some proposals can be found in [4, 5].

Finally we would like to stress, that although the model class is quite specific and maybe not generally applicable, we think that the study of this class is interesting because it is one of the few *"exactly solvable"* multi-agent systems, allowing the study of non-trivial collective optimal behaviour in large distributed systems, both analytically as well as in simulations, and possibly providing insights that might help to develop approximating methods for more general systems.

**Acknowledgments**

This research is part of the Interactive Collaborative Information Systems (ICIS) project, supported by the Dutch Ministry of Economic Affairs, grant BSIK03024.

## A  STOCHASTIC OPTIMAL CONTROL

In this appendix we give a brief derivation of (4), (5) and (6), starting from (3). Details can be found in [4, 5].

The optimal cost-to-go $J$ in a state $x$ at time $t$ is found by minimizing $C(x, t, u(t \to T))$ over all sequences of controls over the time interval $[t, T]$,

$$J(x,t) = \min_{u(t \to T)} C(x, t, u(t \to T)).$$

It satisfies the stochastic Hamilton-Jacobi-Bellman (HJB) equation

$$-\partial_t J = \min_u \left( \frac{1}{2} u^\top R u + V + (b+u)^\top \partial_x J + \frac{1}{2}\mathrm{Tr}(\nu \partial_x^2 J) \right),$$

with boundary condition $J(x, T) = \phi(x)$. The minimization with respect to $u$ yields

$$u = -R^{-1} \partial_x J, \qquad (10)$$

which defines the optimal control. Substituting this control in the HJB equation gives a non-linear equation for $J$. We can remove the non-linearity by using a log transformation: define $\psi(x,t)$ through $J(x,t) = -\lambda \log \psi(x,t)$, with $\lambda$ a constant to be defined, then

$$\begin{aligned}
\frac{1}{2} u^\top R u + u^\top \partial_x J &= -\frac{1}{2}\lambda^2 \psi^{-2}(\partial_x \psi)^\top R^{-1} \partial_x \psi, \\
\frac{1}{2}\mathrm{Tr}(\nu \partial_x^2 J) &= \frac{1}{2}\lambda \psi^{-2}(\partial_x \psi)^\top \nu \partial_x \psi \\
&\quad - \frac{1}{2}\lambda \psi^{-1} \mathrm{Tr}(\nu \partial_x^2 \psi).
\end{aligned}$$

The terms quadratic in $\psi$ vanish if there exists a scalar $\lambda$ such that

$$\nu = \lambda R^{-1}. \qquad (11)$$

In the one dimensional case, such a $\lambda$ can always be found. In the higher dimensional case, this restricts the matrices $R^{-1} \propto \nu$. When (11) is satisfied, the HJB equation becomes

$$\begin{aligned}
\partial_t \psi &= \left( \frac{V}{\lambda} - b^\top \partial_x - \frac{1}{2}\mathrm{Tr}(\nu \partial_x^2) \right) \psi \\
&= -H\psi, \qquad (12)
\end{aligned}$$

where $H$ a linear operator acting on the function $\psi$. Equation (12) must be solved backwards in time with $\psi(x,T) = e^{-\phi(x)/\lambda}$. However, the linearity allows us to reverse the direction of computation, replacing it by a diffusion process, as we will now explain.

The solution to equation (12) is given by

$$\psi(x,t) = \int dy \rho(y, T | x, t) e^{-\phi(y)/\lambda}, \qquad (13)$$

the density $\rho(y, \vartheta | x, t)$ ($t < \vartheta \leq T$) satisfying a forward Fokker-Planck equation

$$\partial_\vartheta \rho(y, \vartheta | x, t) = H^\dagger \rho(y, \vartheta | x, t), \qquad (14)$$

where $H^\dagger$ the Hermitian adjoint of $H$,

$$H^\dagger \rho = -\frac{V}{\lambda}\rho - \partial_y^\top b\rho + \frac{1}{2}\mathrm{Tr}(\nu \partial_y^2 \rho).$$

The potential $V$ in $H^\dagger$ implies an annihilation, the diffusion process is "killed" with a rate $\frac{V}{\lambda} dt$.

Finally we find the optimal cost-to-go from equation (13),

$$J(x,t) = -\lambda \log \int dy \rho(y, T | x, t) e^{-\phi(y)/\lambda}. \qquad (15)$$

## References

[1] C. Boutilier. Planning, learning and coordination in multiagent decision processes. In *TARK*, volume 6, pages 195–210, 1996.

[2] C. Guestrin, D. Koller, and R. Parr. Multiagent planning with factored MDPs. In *NIPS*, volume 14, pages 1523–1530, 2002.

[3] C. Guestrin, S. Venkataraman, and D. Koller. Context-specific multiagent coordination and planning with factored MDPs. In *AAAI*, volume 18, pages 253–259, 2002.

[4] H. J. Kappen. Linear theory for control of nonlinear stochastic systems. *Physical Review Letters*, 95(20):200201, November 2005.

[5] H. J. Kappen. Path integrals and symmetry breaking for optimal control theory. *Journal of statistical mechanics: theory and experiment*, page P11011, November 2005.

[6] R. Stengel. *Optimal Control and Estimation*. Dover, New York, 1993.

[7] S.L. Lauritzen and D.J. Spiegelhalter. Local computations with probabilties on graphical structures and their application to expert systems (with discussion). *J. Royal Statistical Society Series B*, 50:157–224, 1988.